\documentclass[aps,prl,twocolumn,showpacs,superscriptaddress,groupedaddress]{revtex4-1}  % for review and submission
\usepackage{graphicx}  % needed for figures
\usepackage{subfigure}
\usepackage{dcolumn}   % needed for some tables
\usepackage{bm}        % for math
\usepackage{amssymb}   % for math
\usepackage{color} 
\usepackage{epstopdf}
\begin{document}

% The following information is for internal review, please remove them for submission
\widetext
%\leftline{Version xx as of \today}
\leftline{Primary authors: Bruna Amin Gon\c calves, Laura C. Carpi, O. A. Rosso, Mart\'in G. Ravetti, A. P. F. Atman }
%\leftline{To be submitted to PRL}
%\centerline{\em D\O\ INTERNAL DOCUMENT -- NOT FOR PUBLIC DISTRIBUTION}

%\title{Identifying common dynamical patterns in Financial Market crisis}
\title{Quantifying instabilities in Financial Markets}
%\input author_list.tex       % D0 authors (remove the first 3 lines
                             % of this file prior to submission, they
                             % contain a time stamp for the authorlist)
                             % (includes institutions and visitors)

\author{Bruna Amin Gon\c calves}
\affiliation{Programa de P\'os Gradua\c c\~ao em Modelagem Matem\'atica e Computacional, PPGMMC, Centro Federal de Educa\c c\~ao Tecnol\'ogica de Minas Gerais, CEFET-MG. Av. Amazonas, 7675. 30510-000. Belo Horizonte, MG, Brazil}

\author{Laura Carpi}
\affiliation{Departament de F\'{\i}sica i Enginyeria Nuclear, Universitat Polit\`ecnica de Catalunya, 08222 Terrassa, Spain}

\author{Osvaldo A. Rosso}
\affiliation{Departamento de F\'isica, Universidade Federal de Alagoas, 57072-970, Macei\'o, AL, Brazil}
\affiliation{Departamento de Inform\'atica en Salud, Hospital Italiano de Buenos Aires, C1199ABB, Ciudad Aut\'onoma de Buenos Aires, Argentina. }
%\affiliation{Instituto Tecnol\'ogico de Buenos Aires (ITBA), and CONICET, Ciudad Aut\'onoma de Buenos Aires, Argentina}
\affiliation{Facultad de Ingenier\'{\i}a y Ciencias Aplicadas, Universidad de los Andes, Santiago, Chile.}

\author{Mart\'in G. Ravetti}
\affiliation{Departamento de Engenharia de Produ\c c\~ao, Universidade Federal de Minas Gerais, 31270-901, Belo Horizonte, MG, Brazil}

\author{A.P.F. Atman}
\affiliation{Programa de P\'os Gradua\c c\~ao em Modelagem Matem\'atica e Computacional, PPGMMC, Centro Federal de Educa\c c\~ao Tecnol\'ogica de Minas Gerais, CEFET-MG. Av. Amazonas, 7675. 30510-000. Belo Horizonte, MG, Brazil}
\affiliation{Departamento de F\'{\i}sica e Matem\'atica and National Institute of Science and Technology for Complex Systems, CEFET-MG.}
\email[corresponding author ]{ atman@dppg.cefetmg.br}

%\address[Fifth]{Complex Systems Group, Facultad de Ingenier\'{\i}a y Ciencias Aplicadas, Universidad de los Andes, Las Condes, Santiago, Chile.} 

% the following line is for submission, including submission to the arXiv!!
%\hspace{5.2in} \mbox{Fermilab-Pub-04/xxx-E}

\date{\today}

\begin{abstract}
 Financial global crisis has devastating impacts to economies since early XX century and continues to impose increasing collateral damages for governments, enterprises, and society in general. Up to now, all efforts to obtain efficient methods to predict these events have been disappointing. However, the quest for a robust estimator of the degree of the market efficiency, or even, a crisis predictor, is still one of the most studied subjects in the field. 
We present here an original contribution that combines Information Theory with graph concepts, to study the return rate series of 32 global trade markets. Specifically, we propose a very simple quantifier that shows to be highly correlated with global financial instability periods, being also a good estimator of the market crisis risk and market resilience. We show that this estimator displays striking results when applied to countries that played central roles during the last major global market crisis. 
The simplicity and effectiveness of our quantifier allow us to anticipate its use in a wide range of disciplines.
\end{abstract}

\maketitle

The Wall Street 1929 crash made modern society aware of one of the most destructive events of capitalism, the so-called \emph{Global Market Crisis}. With devastating consequences, these crisis led enterprises, financial corporations and even governments to broke up causing poverty, panic and death all around the world. A crescent combination of efforts has been dispensed to tame the task of unveiling the driving mechanisms of market dynamics \cite{bouchaud2003,kiyono2006}. Recently, the treatment of Financial Markets (FM) as paradigmatic examples of complex systems~\cite{atman2012,Battiston2016,Battiston2016A} led to the application of network theory~\cite{Battiston2010,stefan2015}. This new approach has focused on the analysis and modeling of the dynamic of these systems, to understand better the underlying mechanisms driving the emergence of instability periods~\cite{Bardoscia2017}. These network tools have the potential to describe the current interconnected financial system, giving new perspectives to follow up and management of unexpected events~\cite{Stanley2008,zunino2008,cajueiro2009}.

This work proposes a novel quantifier defined as the ratio of the Shannon entropy by the Fisher Information measure --  the $\mathcal {SF}$ index  -- computed over probability distributions (PDFs) extracted from return rate graphs of stock markets from different countries. Return rates graphs are constructed from return rate time series by considering the difference of the amplitude's values of points that are connected through a ``visibility'' criteria, obtaining in this way, a weighted graph. A weighted probability distribution then represents the topology of the graph $P_w$, in which the weight is a real value proportional to the amplitude difference between the two connected points, aggregating temporal and spatial features of the time series~\cite{Goncalves2016}. The Visibility Graph algorithm (VG) \cite{lacasa2008}, has been widely used and successfully applied to study many different systems (see Aragoneses et al.~\cite{Aragoneses2016} and references therein). However, to the best of our knowledge, despite some efforts to apply VG to financial data \cite{mutua2015,yang2009}, there is still no work dealing with the estimation of market volatility. The VG allows the use of well-known Information Theory quantifiers to analyze features of the system, which are not simple, or even impossible to be assessed with traditional methods \cite{Ravetti2014, Schieber2017}. Details of the construction of VG and weight distribution $P_w$ are discussed in the Supplementary Information (SI). 
The $\mathcal {SF}$ index is a  very simple and practical tool to perform this task since it considers well-known entropic based information measures of graph properties. The combination of the Shannon entropy $\mathcal {S}$~\cite{Shannon1948} and Fisher Information $\mathcal {F}$~\cite{Fisher1922} in the $\mathcal {SF}$ index is highly efficient in characterizing the market's dynamic, since it systematically increases during crisis periods. 

\begin{figure*}
 \includegraphics[width=0.8\linewidth]{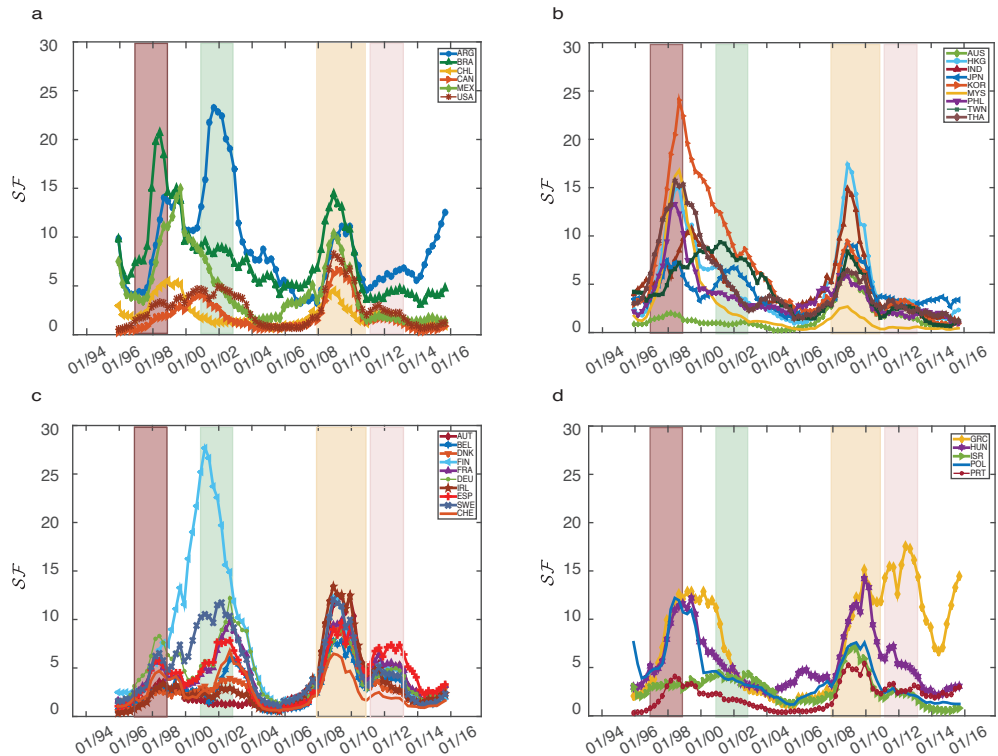}
\caption{\label{fig:SF} These panels show the $\mathcal {SF}$ index calculated from weighted visibility graphs built considering a time span window of two years and three months gap between windows. In panel {\bf a} are showed results for American countries, panel {\bf b} for Asiatic markets, and panels {\bf c} and {\bf d}, for European more developed, and less developed countries, respectively.}
 \end{figure*}

$\mathcal {SF}$ combines global and local aspects of the PDF's associated to the studied system. $\mathcal {S}$ is a measure of global character that it is not too sensitive to local changes in the distribution.
$\mathcal S [P] = 0$, means that there is a single state active in the system and, thus, it is possible to predict the behavior of the system with maximal knowledge. 
%Considering the degree distribution of a graph, $P_k$, $S[P_k]=0$ would correspond to a regular network. Otherwise, for a uniform distribution, $P_U =\{p_i = 1 / N~ \forall i=1,\cdots,N\}$, then, $S[P_U] = S_{max} = \log_b (N)$, value that can be used to normalize $\mathcal S [P] $, as $H[P]~=~S[P]/ S_{max}$.  In the case of the weight distribution $P_w$, $N$ is equal to the number of bins used to build the frequency count. 

The maximal $\mathcal{F}$ value, $\mathcal{F}_{max}$, is reached when the knowledge about the system is complete while it vanishes when the information about the system is minimal, in a directly opposite sense compared with Shannon entropy. Thus, since the $\mathcal{SF}$ index is obtained by dividing the value of Shannon entropy by the Fisher Information measure, it is significantly more sensitive to any change in the degree of knowledge of the system, either local as global ones.  $\mathcal{S}$ and $\mathcal{F}$ are formally defined in SI (Supplemental Information).

To apply the techniques presented above into return rate time series we need a standardization of the collected data from different countries. First, the weekends are not considered once stock markets are closed, and there is no record in the time series; second, each country has a different number of holidays during the year. Thus, we apply a linear interpolation in the original time series to fill up these gaps. The period considered is comprised between January 1995 to May 2016, collected from Bloomberg$^{TM}$ system. We examine the daily closing prices to calculate the return rates for the 32 countries listed in SI.

\begin{figure}
\includegraphics[width=1\linewidth]{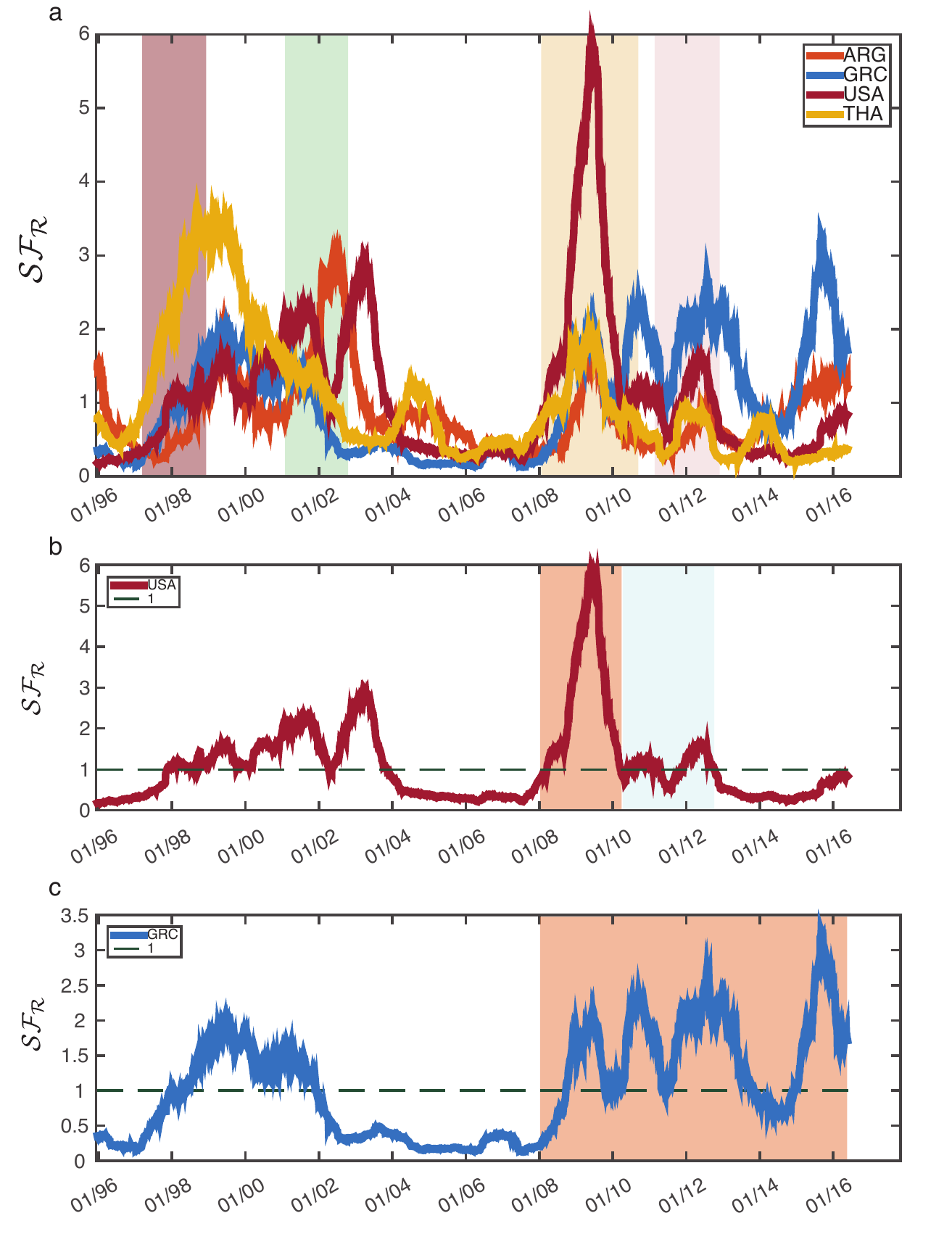}
 \caption{\label{fig:estimator}  This plot shows the $\mathcal {SF_R}$  \emph{risk estimator} for graphs constructed considering the 12 previous months to the corresponding day. Panel {\bf a} shows that this estimator can identify which countries are most affected during crisis periods, as one can observe in the plot for the so-called $1998$ Asiatic crisis, the $2000$'s Argentina default crisis, $2008-9$ subprime mortgage crisis, which have implied particularly the USA, and the recent Greek government recurrent crisis. Panels {\bf b} and {\bf c}, enhance the behavior of USA and Greece, depicting an instability period in orange. In light blue we depicted a post-instability period for USA, with $\mathcal {SF_R}$ values oscillating closely to 1, followed by a more stable period. Although with a more intense crisis, reaching $\mathcal {SF_R}=6$, USA showed a strong resilience with a rapid recovery. Greece does not seem to overcome its instability period yet. 
}
\end{figure}

Several different parameters settings were tested to improved the performance of the method, see Figures S2. We present here two sets of results considering different time span windows and gaps. The first ensemble uses a time span window of $504$ points (two years), lagged by $63$ values (three months). For each graph, we calculate the Shannon entropy and Fisher Information measure considering a fixed size of $50$ bins histograms of the weight distribution $P_w$. Fig. \ref{fig:SF} depicts these results showing the temporal evolution of the index for each country. The first striking feature that can be immediately inferred is that all countries present a systematic behavior, displaying higher values of $\mathcal {SF}$ index during crises periods, and lower values during inter-crises periods. Four terms of crises are enhanced in those panels: the Asiatic crisis (1997-98); Argentine crisis and dot-com bubble (1999-2001); the USA subprime mortgage crisis (2007-09), the more recent worldwide bank crisis, Icelandic financial crisis (2008-2012) and Greek government debt crisis (2009-2012). Another period which led to global instabilities was the terrorist attack on Twin Towers ($9/11/2001$) which coincided with the acutest period of the Argentina crisis (2001-2002). It is remarkable that, in each crisis period, the higher $\mathcal {SF}$ index value displayed in the panels corresponds exactly to the country most affected during that period. As an exception, we point out  Finland, which shows a prominent peak just before the Argentine crisis of 1999. It worth to say that Finland has experienced a severe banking crisis in early of 1990's which turned the Finnish market unstable at that time.

A comparison among countries belonging to the different blocks evinces that South American markets display higher values of $\mathcal {SF}$ index, especially during a global crisis. Particularly, Argentine $\mathcal {SF}$ index highest peak coincides with the worst crisis moment in this country \cite{BOSCHI}. In Brazil, the most significant peak is closely related to the Asiatic crisis, corroborating the hypothesis that Brazil was the most affected due to its strong relation with Asiatic developing countries \cite{HARVEL}. Considered the most stable South American country, Chile market displays $\mathcal {SF}$ index values consistently lower compared to other South American countries, again confirming the robustness of the $\mathcal {SF}$ index to quantify market efficiency. This observation also holds for North American countries which displayed the highest values during the USA recession period due to the subprime mortgage crisis, except for Mexico, mostly affected by the Asiatic crisis. Those observations confirm the strong influence of Asiatic crisis in emerging markets as Mexico and Brazil \cite{GRAHAM, KAMINSKY}, compared with the collateral effects in developed countries. A sudden and localized increasing of the USA $\mathcal {SF}$ index curve identifies the $9/11$ terrorist attack.  Marginal effects of this case are seen in the plots of Japan and other developed European countries.

Another remarkable example of the accuracy of the $\mathcal{SF}$ index to quantify market efficiency is observed in the panel of Asiatic countries. Firstly, the curves in Fig. \ref{fig:SF}-b displays sharp peaks exactly during the crisis periods. Secondly, we observe a strong temporal correlation between the time step when the $\mathcal {SF}$ index attains its maximum and the crisis date. For example, the first country implied in the crisis, analyzing the panel \ref{fig:SF}-b is Thailand, and we learn from literature that this country first experienced a substantial currency depreciation \cite{KHALID}. Next, one can observed the increasing of $\mathcal{SF}$ index in India, The Philippines, Honk Kong, Indonesia, etc., reflecting the crisis infection in the Asian markets. This propagation affected then American developing countries -- panel \ref{fig:SF}-a, but have only limited influence in Europe more developed countries -- panel \ref{fig:SF}-c --  both features which reported in literature \cite{Corsetti, CHOWDHRY}. The interdependency of markets is also depicted from the $\mathcal{SF}$ index comportment, as for instance the strong bonds between Israel and USA markets \cite{GRAHAM}, or Honk Kong and the developed markets during the USA subprime crisis \cite{LIM, LOH}. 

European developed countries display higher peaks in $\mathcal{SF}$ index curve correlated with the USA subprime mortgage crisis, but we also notice a significant instability associated with the $9/11$ terrorist attack, corroborating the well-known bond between developed markets,  including Japan \cite{DIMPFL, MOLLAH}. One can observe that the less developed European countries -- Portugal, Poland, Hungary, Greece -- show a higher correlation with the Asiatic crisis (1997-1998), and almost none with the $9/11$.

  A different analysis is performed constructing daily VGs in which it is considered the information of the preceding year. A $\mathcal{SF}$ \emph{risk estimator} ($\mathcal{SF_R}$ ), is defined as the ratio of the $\mathcal{SF}$ index over the corresponding interval, and the $\mathcal{SF}$ index considering the entire time series.  We use the $\mathcal{SF_R}$ as an estimator of the market efficiency, and a quantifier of the risk of a crisis to occur (see Figure~\ref{fig:estimator}). 

The ability of $\mathcal{SF_R}$ to evince which market is the most implied in the crisis is remarkable. It is worth noticing that the value of the estimator starts to increase before the assumed period of crisis, reinforcing the role of this quantifier as an earlier signal of a crisis in a particular market. In this way, every time that $\mathcal{SF_R}$ exceeds the unity in a sequence of consecutive points, it is AN indicator that a crisis can be underway. 

This measure turned out to be a very robust indicator of the market efficiency, as shown in Fig. \ref{fig:estimator}. We believe that $\mathcal{SF_R}$ could be a useful tool to estimate the risk of a given market to experience a crisis, or simply, a measure of the efficiency degree of a market \cite{cajueiro2004rank}.

\begin{figure*}
 \includegraphics[width=1\linewidth]{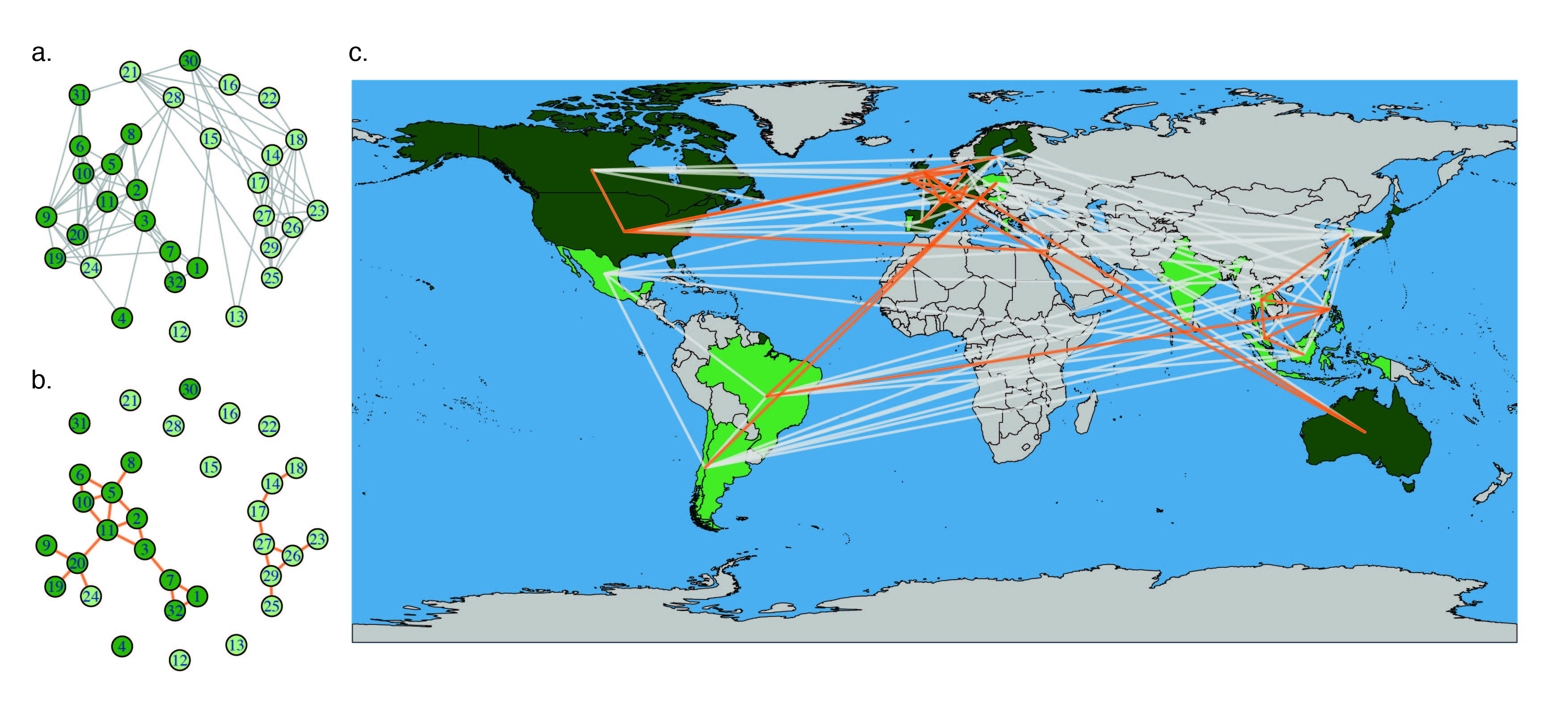}
\caption{\label{fig:maps} { Global market networks constructed by considering different threshold values of pair correlation between countries $\mathcal{SF}$ index sequences. Subplots (a) and (b) depict the networks for thresholds 0.8 and 0.9, respectively. Subplot (c) depicts the connections between countries, represented in a global map, for thresholds 0.8 (white) and 0.9 (orange). The  countries are listed in Section SI.}}
 \end{figure*}

A set of graphs, to study the relation between countries, is constructed considering the Pearson correlation of their $\mathcal{SF_R}$ curves and a given threshold. For correlations above the threshold, is created a link between the countries. As the threshold value increases, it is possible to observe the emergence of clusters of countries which correspond to the commercial blocks, as the Eurozone or the Asiatic countries. Figure \ref{fig:maps} depicts the global networks constructed considering thresholds $0.8$ and $0.9$, with a few exceptions.

It is interesting to observe that Hong Kong links the two clusters, well-known fact in the financial community but hardly be quantified. Another impressive result is that Argentina and Greece appear isolated in this graph. Another solitaire countries are Portugal, linked only with Austria, and Finland, connected with Taiwan and Sweden. Notice that Sweden also experienced bank crisis during the late 1990's just before the Asiatic crises. For a threshold value of 0.9, most of the countries are isolated, and remains only a backbone for emergent and developed countries, with Israel (emerging) placed at the developed ones. Graphs and connectivity maps for thresholds values $0.6$ and $0.7$ can be found in the SI.

 We present a novel estimator to analyze a stochastic time series able to characterize either the local volatility of the series as well its global information entropy properties. We applied the technique to analyze the return rate time series of several countries and observed that the estimator displays a remarkable ability to identify instability periods as well to quantify market efficiency. This striking success of the methodology to analyze financial data should incentive analogous applications in a vast range of areas where signal analysis plays a central role, as biomedical signals, climate indexes among others.
 
\bigskip

{\bf Acknowledgements} 
We are indebted to Professor Felipe D. Paiva, from DCSA - Department of Applied Social Science -  that graciously have granted access to the Bloomberg$^{TM}$ system. M.G.R and A.P.F.A. acknowledge support from CNPq and FAPEMIG (Brazil). B.A.G. thanks to Brazilian agency CAPES by post-doc grant. O.A.R acknowledges support from CONICET (Argentina).

\bibliographystyle{apsrev4-1}

%\bibliographystyle{naturemag}
%\bibliography{Letter_06_04_M}

%\section*{References}
%

\end{document}